\documentclass[preprint]{aastex}
\def\simgr{\,\hbox{\hbox{$ > $}\kern -0.8em \lower 1.0ex\hbox{$\sim$}}\,}
\def\simle{\,\hbox{\hbox{$ < $}\kern -0.8em \lower 1.0ex\hbox{$\sim$}}\,}

\shortauthors{THORSTENSEN, PETERS, AND SKINNER}
\shorttitle{Longer Period CVs}
\begin{document}
\title{Optical Studies of Twenty Longer-Period Cataclysmic Binaries
\footnote{Based on observations obtained at the MDM Observatory, operated by
Dartmouth College, Columbia University, Ohio State University, 
Ohio University, and the University of Michigan.}
}

\author{John R. Thorstensen, Christopher S. Peters, and Julie N. Skinner}
\affil{Department of Physics and Astronomy\\
6127 Wilder Laboratory, Dartmouth College\\
Hanover, NH 03755-3528}

\begin{abstract}
We obtained time-series radial velocity spectroscopy of twenty
cataclysmic variable stars, with the aim of determining orbital periods $P_{\rm orb}$.
All of the stars reported here prove to have 
$P_{\rm orb} > 3.5$ h.  For sixteen of the stars, these are the
first available period determinations, and for the remaining
four (V709 Cas, AF Cam, V1062 Tau, and RX J2133+51) we use new
observations to improve
the accuracy of previously-published periods. 
Most of the targets are dwarf novae,
without notable idiosyncracies.  Of the remainder, three (V709 Cas, 
V1062 Tau, and RX J2133+51) are intermediate polars (DQ Her stars);
one (IPHAS 0345) is a secondary-dominated system without known
outbursts, similar to LY UMa; one (V1059 Sgr) is an old nova; 
and two others (V478 Her and V1082 Sgr) are long-period novalike
variables.
The stars with new periods are 
IPHAS 0345 (0.314 d); 
V344 Ori (0.234 d); 
VZ Sex (0.149 d);
NSVS 1057+09 (0.376 d); 
V478 Her (0.629 d);
V1059 Sgr (0.286 d);
V1082 Sgr (0.868 d);
FO Aql (0.217 d);
V587 Lyr (0.275 d);
V792 Cyg (0.297 d);
V795 Cyg (0.181 d);
V811 Cyg (0.157 d);
V542 Cyg (0.182 d);
PQ Aql (0.247 d);
V516 Cyg (0.171 d);
and 
VZ Aqr(0.161 d). 
Noteworthy results on individual stars are as follows.
We see no indication of the underlying white
dwarf star in V709 Cas, as has been previously claimed; based
on the non-detection of the secondary star, we argue that 
the system is farther away that had been thought and the white
dwarf contribution is probably negligible.  V478 Her had been 
classified as an SU UMa-type dwarf nova, but this is 
incompatible with the long orbital period we find.
We report the first secondary-star velocity curve for
V1062 Tau.  
In V542 Cyg, we find a late-type contribution that remains
stationary in radial velocity, yet the system 
is unresolved in a direct image, suggesting that 
it is a hierarchical triple system.

\end{abstract}

\keywords{keywords: stars}

\section{Introduction}
Cataclysmic variables (CVs) are close binary star systems in which a white 
dwarf primary star accretes matter from the Roche lobe overflow of 
a less-compact secondary star.  \citet{war95} gives an 
overview of the field.   

The orbital period, $P_{\rm orb}$, of a CV is its most easily 
measurable fundamental property.  Mass transfer in CVs is thought to
be driven by the slow loss of orbital angular momentum. During
most of a CV's life, this leads to gradual shrinkage of the orbit and
consequent tightening of secondary's Roche lobe.  The
orbital period therefore gives a rough idea of a system's 
evolutionary state.  Once an orbital period is known, the 
Roche-filling criterion sets strong constraints on the 
secondary's radius, which in favorable circumstances can be
exploited to yield a distance estimate. 

CVs with high orbital inclinations $i$ will eclipse, which makes
orbital period determination straightforward and also makes
it possible to estimate component masses and the like.
Non-eclipsing systems sometimes show periodic photometric 
modulations, but these do not always appear at 
$P_{\rm orb}$; other phenomena such as 
superhumps \citep{patterson05} can mimic orbital modulations.  
Radial velocity spectroscopy is therefore the
most reliable way of obtaining $P_{\rm orb}$ for a
non-eclipsing CV
\footnote{In rare cases, a radial-velocity modulation may 
not be at $P_{\rm orb}$; \citet{araujo05} discuss a
particularly interesting example.}.   For this reason 
we have therefore been
collecting radial-velocity time series for a large number
of CVs, with the aim of determining $P_{\rm orb}$. 

In this paper we collect together observations and analyses
of twenty CVs, all of which proved to have periods
well longward of the so-called period gap, a range of
$P_{\rm orb}$ from 
roughly 2 to 3 hours in which relatively few CVs are found
\citep{gapstillthere}.
Table~\ref{tab:star_info} gives basic information about the
stars studied here.  The periods of most of these stars
are published here for the first time; a few others are 
included because we have improved the accuracy of the period.
We did not use any particular astrophysical criterion to 
select the present sample, but rather observed these stars
because period determinations appeared feasible.

Section 2 describes the 
observations and analysis, and \S 3 gives results for the
individual objects.  

\section{Techniques}

Our spectroscopic observations are from the 2.4 m Hiltner or 1.3 m 
McGraw-Hill telescopes at MDM 
Observatory in Kitt Peak, Arizona.  At the Hiltner, we 
used the 
`modular' spectrograph, a 600 line mm$^{-1}$ grating, a 1-arcsec slit, and 
a SITe 2048$^2$ CCD detector.  Most of the spectra cover from 
4210 to 7560 \AA\ (vignetting severely 
toward the ends), and have typical resolution of 3.5 
\AA\ FWHM.  On the 1.3m, we used the Mark III spectrograph, a SITe 
1024$^2$ CCD detector, a 2-arcsecond slit, and achieved typical 
resolution of 4-5 \AA\ FWHM from 4645 to 6960 \AA.  On earlier
observing runs, we took spectra 
of comparison lamps whenever the telescope
was moved to achieve accurate wavelength calibration, but on later
runs we used the night-sky spectrum to shift the zero point
of a master wavelength calibration obtained during twilight.
When the sky was clear, we observed bright early-type stars to 
derive corrections for the telluric absorption features, and flux 
standards to calibrate the instrument response.  Our spectrophotometry
suffers from uncalibratable errors due to losses at the spectrograph
slit, but experience suggests that these typically amount to 
$\sim 0.2$ mag.  \citet[hereafter TFT04]{thfe04} 
discuss the observational protocols in more detail.

\subsection{Radial Velocity Measurement and Period Determination}

We measured centroids for the emission lines using convolution 
techniques explained by \citet{sy80} and \citet{shafter83}, in which
the line profile is convolved with an antisymmetric function and the
zero of the convolution is taken as the line center.  For the 
most part we used a convolution function consisting of a positive
Gaussian and a negative Gaussian separated by an adjustable amount. 
We usually set the separation of the Gaussians to maximize 
sensitivity to the steep sides of the line profile.

The Roche geometry dictates that longer-period CVs will have 
larger and generally more massive secondary stars than shorter-period
CVs, so their secondary stars often contribute significantly to the
total light.  When possible, we measured absorption line velocities using the 
cross-correlation radial velocity package {\it xcsao} 
\citep{kurtzmink}.  In most cases the template spectrum was a composite of 
76 observations of late-type IAU velocity standard stars, which had
been shifted individually to zero velocity before averaging; for
later-type secondaries we used a similar template constructed
from M dwarfs.
The spectral region used in the cross-correlation was either 
6000 to 6500 \AA, or a window from 5100 to 6500 \AA, excluding
the complicated region near HeI $\lambda5876$ and NaD.
Uncertainties, based on the $R$-statistic of \citet{tondav}, were typically 
on the order of 20 km s$^{-1}$.  We omitted unphysically large velocities 
(which result when the `wrong' cross-correlation peak is selected) and 
velocities with large estimated uncertainties.

We analyzed the velocity time series with a `residualgram' period-search
algorithm \citep{tpst}   
which fits general least-squares sinusoids of the form 
$$v(t) = A\cos(\omega t) + B\sin(\omega t) + C$$ to the time series over
a densely-spaced grid of trial frequencies
$\omega$.  The results are presented as plots of 1/$\chi^2$, where
$\chi^2 = \sum[(v_i - v(t_i))/\sigma_i]^2$; here $v_i$ is a
measured velocity, $v(t_i)$ is the best-fitting sinusoid
evaluated at the corresponding time $t_i$, and $\sigma_i$ is the formal
uncertainty of the velocity, obtained by propagating the estimated 
counting-statistics uncertainties of the individual wavelength channels. 
Ideally, a graph of 
$1/\chi^2$ as a function of $\omega$ shows a strong spike at the 
orbital frequency, but the imperfect 
sampling of the time series inevitably 
results in aliasing. 
We used a Monte Carlo algorithm developed by 
\citet{thfreed85} to assess the confidence with which the highest peak in the 
periodogram can be identified with the true orbital 
frequency.  Once we were satisfied that the period was correct, we fit the 
time-series with sinusoids of the 
form $$v(t) = \gamma + K\sin[2 \pi(t - T_0)/P]$$ using a hybrid linear 
least-square algorithm\footnote{Emission- and 
absorption-line velocity amplitudes are denoted as $K_1$ and $K_2$, 
respectively}.  This procedure is 
described in detail in TFT04.  
For those CVs with absorption-line radial velocity solutions,
we shifted the individual spectra into the absorption-line
rest frame before averaging them.

\subsection{Secondary Stars and Spectroscopic Distances}

As noted earlier, the secondary-star spectrum is often
detectable in longer-period CVs.  The absorption-line radial 
velocities should track the orbital motion of the 
secondary, though the radial velocities may be affected
by irradiation and other effects \citep{davey92}.
However, a late-type contribution need not 
arise from the secondary; if the late-type spectrum is nearly
fixed in velocity, it may come from an unrelated field star,
or the outer member of a hierarchical triple;
This appears be the case for V542 Cyg, discussed further below.

\citet{peters05} and TFT04 discuss our spectral decomposition 
technique and the inferences based on the secondary 
star, which we summarize briefly here.  
To estimate the secondary star's spectral type and 
its contribution to the light, we scale spectra of cool
spectral-type standard stars, subtract them from the
target spectrum, and look for the best possible 
cancellation of the late-type features.  This yields
estimates of the spectral type of the secondary and
its contribution to the total light.  The spectral type
constrains the secondary's surface brightness, and the 
orbital period (together with the condition that the
secondary fills its Roche critical lobe) tightly constrains
the secondary's radius $R_2$.  The precise value of 
$R_2$ does depend weakly on the secondary's mass, $M_2$;
to compute a realistic range for $R_2$, we use the
evolutionary calculations of \citet{bk00} to 
guide the selection of a range for $M_2$.
The estimates of the radius and surface brightness
together yield the secondary's absolute magnitude, which 
in turn gives the distance, without any assumption that the
secondary is a typical main-sequence star.  Where 
applicable, we estimate the visual extinction 
using the \citet{schlegel98} maps; these give the 
extinction to the edge of the Galaxy, so we use their
tabulated $E(B-V)$ as an upper limit.   

\subsection{Direct Imaging and Photometry}

For most of the stars studied here, we have collected direct
images using the Hiltner 2.4m telescope and a SITe 2048$^2$
CCD detector.  Most of the celestial positions in Table~\ref{tab:star_info}
were measured from these images.  In a few cases we
give magnitudes based on $UBVI$ or $VI$ image
sets, which were reduced using observations of 
standard stars from \citet{landolt92}.

\section{The Individual Stars}

\begin{figure}
\epsscale{0.95}
\plotone{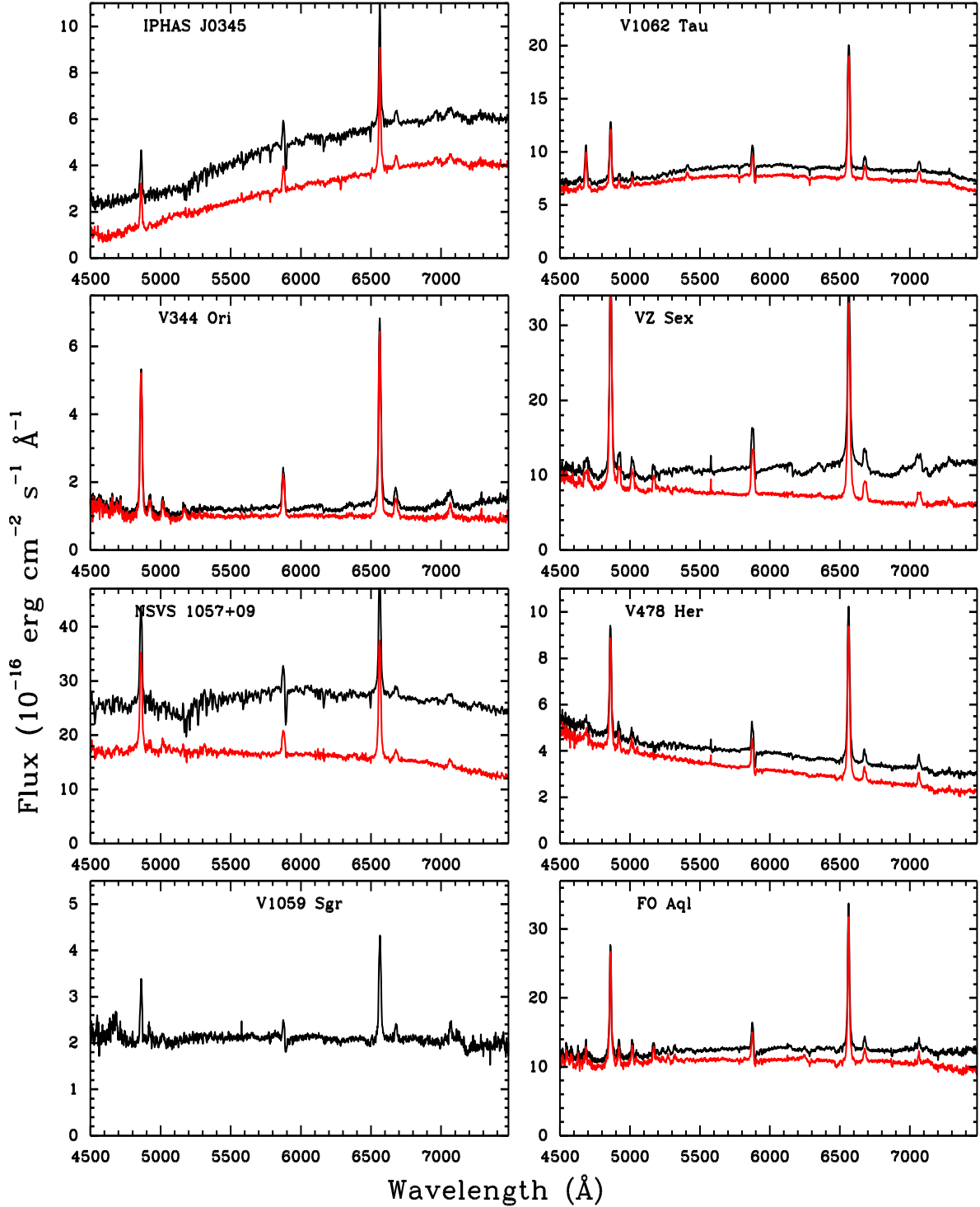}
\caption{Mean spectra of CVs.  In cases where two traces
are given, the lower trace (red in the on-line edition) is
the mean spectrum after a scaled, late-type spectrum
has been subtracted.  V709 Cas is shown separately in
Fig.~\ref{fig:v709casspec}. 
}
\label{fig:spectra1}
\end{figure}

\begin{figure}
\epsscale{0.95}
\plotone{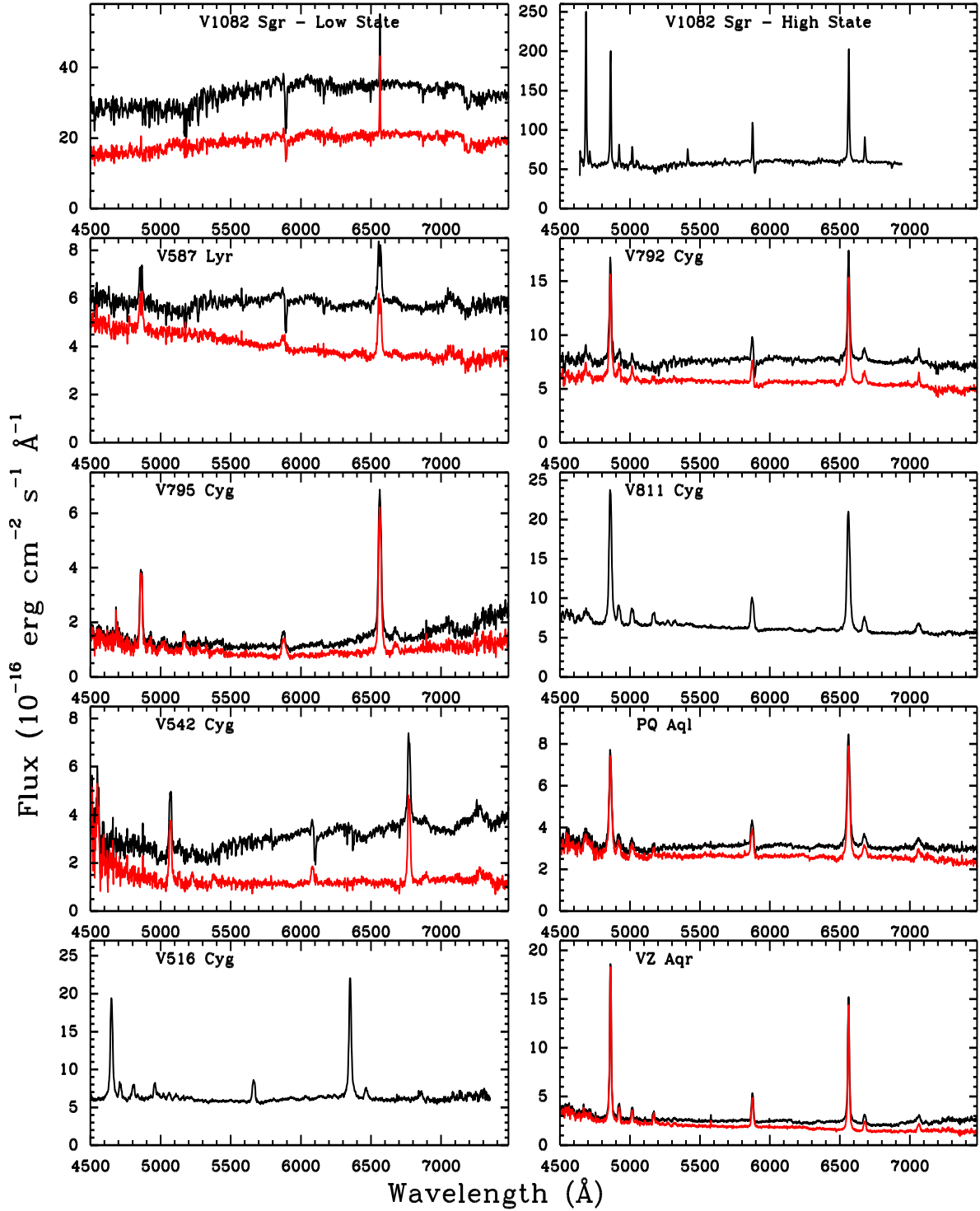}
\caption{Same as Fig.~\ref{fig:spectra1}, for the remainder
of the stars, except for RX J2133.7+51057, which is not shown.
V1082 Sgr is shown in its high and low states.
}
\label{fig:spectra2}
\end{figure}

Our results for the individual CVs are summarized in tables and 
figures.  
Figs.\ref{fig:spectra1} and \ref{fig:spectra2} show 
the mean spectra of most of the stars (excluding a few
spectra of dwarf novae in outburst).
A small sample 
of radial velocity measurements included in the analysis are listed 
in Table~\ref{tab:velocities}\footnote{The complete table can 
be found in the electronic edition.}.  
Figs.~\ref{fig:pgrams1} and \ref{fig:pgrams2}
give periodograms 
of the velocity time series, computed as described above.
Table~\ref{tab:parameters} lists parameters of the best-fit sinusoids to the 
velocity time series.  For all parameters, the number in parentheses 
represents the uncertainty in the last digit.  
Finally, 
Figs.~\ref{fig:folpl1}, \ref{fig:folpl2},  and \ref{fig:folpl3} 
show folded radial 
velocity curves; when both absorption- and emission-line 
velocities are available, the radial velocity curves are folded about the 
weighted mean period.  %

\subsection{V709 Cas} 

V709 Cas is a DQ Her star, or intermediate polar (IP), with a pulsation
period near 313 s.  \citet{bonnetv709} found $P_{\rm orb} = 0.2225 \pm 
0.0002$ d from a radial-velocity study.  They also detected absorption
wings around the higher Balmer lines, which they interpreted as arising
from the white dwarf photosphere.  Our H$\alpha$ emission-line radial 
velocities were taken in the interval from 1996 to 2005, and define a 
unique cycle count over the entire time span.  The period we find, 
0.2222033(7) d, or 5.33288 hr, agrees within
the uncertainties with the \citet{bonnetv709} period, but is more
precise because of the longer time base.

Fig.~\ref{fig:v709casspec} shows the average spectrum from 1999 October;
the flux level in this spectrum corresponds to $V = 15.0$.  
While the spectrum does not cover the higher Balmer lines well, 
we do not see absorption wings around H$\beta$ comparable to those
found by \citet{bonnetv709}.  
We examined the summed spectra from other observing runs, 
and none of these showed Balmer 
absorption either.  We therefore cannot confirm the white
dwarf detection claimed by \citet{bonnetv709}.  

After some consideration, we are skeptical that the Balmer absorption
wings detected by \citet{bonnetv709} arise from the white dwarf photosphere.
They get a good fit to the absorption using a 
$\sim 23000$ K white dwarf model atmosphere.  At $\log g = 8$ 
(in cgs units), a white dwarf at this $T_{\rm eff}$ should have 
$M_V \sim 10$ \citep{bergeron}.  If the system's luminosity is low enough for 
such a faint white dwarf to be seen, then where is the secondary?
\citet{kniggedonor} compiled a semi-empirical `donor sequence'
for CVs, and at the 5.3 hr orbital period of V709 Cas, 
this predicts the spectral class of the secondary to be 
near M2, and the absolute magnitude $M_V$ to be near 9.
For their interpretation to be correct, the flux contributed by
the white dwarf puts the system at $\sim 230$ pc, or
$m - M \sim 6.8$; the $M_V \sim 9$ secondary should 
therefore have $V \sim 16$, and contribute very substantially
to the spectrum, but we do not detect the secondary's contribution
in our long cumulative exposures.  We do see NaD features, with
each of the D lines having an equivalent width of $\sim 0.5$ \AA ,
but these are sharp and motionless, indicating an interstellar
origin.  We feel that the balance of the evidence suggests 
that the line wings are an intermittent feature of the 
optically thick disk, and that the system is more luminous
and distant than in the \citet{bonnetv709} interpretation.

\subsection{AF Cam}

\citet{thorstensen01} reported spectroscopy of AF Cam.  They
detected the K-type secondary star, and radial velocity
variability in both emission and absorption lines at a 
weighted mean period of $0.324 \pm 0.001$ d.  We have since
observed the star occasionally with the aim of refining 
$P_{\rm orb}$ and deriving an ephemeris.  The period and
epoch given in Table~\ref{tab:parameters} is based on an
unambiguous cycle count over the interval from 2000 to 2007.
\citet{thorstensen01} show a spectrum and analyze the 
secondary star's contribution.

\subsection{IPHAS J034511.59+533514.5}

\citet{iphascvs} discovered this CV as part of the Isaac Newton Telescope 
Photometric H$\alpha$ Survey of the northern Galactic Plane,
The strong K-type contribution evident in their spectrum, and the
relatively bright magnitude ($r' = 16.0$) made this a tempting target
for an absorption-line velocity study, which we undertook starting in 
2008 January.   The expected modulation did prove to be present
in both the H$\alpha$ emission velocities and the absorption 
velocities, with $P_{\rm orb}$ near 7.5 hr.  Both time series
favor a period near 0.31390 d.  An alias period near 0.31609
d, reflecting an alternate choice of cycle count between
observing runs, is much less likely.

The secondary spectrum is best matched by a K5.5 $\pm$ 1.5 star,
but the continuum that remains after the secondary star is
removed is unusually red, suggesting that the reddening is 
substantial.  The Galactic latitude is low, $b = -0^{\circ}.95$,
and the \citet{schlegel98} maps give the reddening $E(B-V) = 2.4$ 
mag at this location.  We experimented with de-reddening by 
various amounts.  With reddenings $E(B-V) \sim 2.0$,
the Balmer decrement became implausibly steep 
(H$\beta$ stronger than H$\alpha$), while at small reddenings the continuum
remained too red.  The best values appeared to be in the 
range $E(B-V) = 0.9 \pm 0.4$.  The procedure outlined
earlier then led to the inferences in Table \ref{tab:inferences}.
The distance estimate is very rough, mostly because of the 
large uncertainty in the reddening.

Sinusoidal fits to the emission and absorption velocities show
them to be $0.56 \pm 0.02$ cycles out of phase.  In 
double-lined spectroscopic binaries, the ratio $K_2/K_1$ gives an
estimate of the mass ratio, but in order for this to be
accurate, the velocities must trace the center-of-mass
motion of the two stars.  If this is the case, the phase
lag between emission and absorption must be 0.5 cycles --
this is a necessary, but not sufficient, condition for the 
estimate to be trustworthy.  In the present case, this condition
is marginally fulfilled -- the discrepancy is formally three 
standard deviations.   
If we assume the $K$-velocities 
do reflect the center-of-mass motion, we find $q = M_2 / M_1 = 
0.83 \pm 0.13$.  Combining this with a broadly typical 
white dwarf mass of 0.7 M$_\odot$ yields a secondary 
mass near 0.55 M$_\odot$, similar to what the 
\citet{bk00} models would suggest at this period and 
spectral type.  

While these mass calculations are somewhat fanciful, the 
modest value of $K_2$ does firmly constrain the orbital
inclination $i$ to be fairly low.
Even we take $M_1$ to be a very low
$0.4$ M$_{\odot}$ and $q$ to be 0.6 -- pushing 
both quantities to values that increase $i$ -- the 
inclination is only $\sim$ 25 degrees.  The apparent
lack of eclipses is therefore not unexpected.

The \citet{bk00} evolutionary models that best match this system
are those in which mass transfer begins after 
hydrogen has been substantially depleted in the core.
The spectral type of the secondary is similar to that in 
other CVs with similar orbital periods \citep{kniggedonor}.

No dwarf nova outbursts have been reported from this system.
In many respects it resembles LY UMa (= CW 1045+525), a 0.271-d
emission-line binary with a strong late-K contribution and no 
reported outbursts \citep{tappert01}.

%
%
\subsection{V1062 Tau}

V1062 Tau is the optical counterpart of the X-ray source 1H0459+246, 
which was discovered in the HEAO-2 Large Area Sky Survey \citep{wood84}.
\citet{remillard94} found an X-ray modulation near 62 minutes,
and classified the object as an intermediate polar, or DQ Her star.
The 62-minute X-ray modulation was confirmed and refined by \citet{hellier02}.
\citet{remillard94} also published an optical spectrum, and noted 
the features of a CV and a K star.  They obtained  
$V$ and $I$ time series photometry that showed ellipsoidal variation 
at a period of $9.95 \pm 0.07$ h.  \citet{lipkin04} monitored
V1062 Tau photometrically, and refined the orbital period to 
$9.9802 \pm 0.0006$ h\footnote{
Their paper quotes this figure as 9.9082 h, but Y. Lipkin (private
communication) confirms that two digits were mistakenly transposed.  
}.  They also detected
the 62-minute period, and the beat between this and the orbit.
Their data showed two short, low-amplitude outbursts, with
$\Delta I \sim 1.2$ mag.

For the most part, our spectra of V1062 Tau appeared similar to 
the average shown in Fig.~\ref{fig:spectra1}, which in turn 
is generally similar to that shown by \citet{remillard94} (their
Figure 3). The excitation is high; HeII $\lambda 4686$ emission is 
similar in strength to 
H$\beta$, and our long cumulative exposure also shows 
emission at HeII $\lambda 5411$.  The \citet{schlegel98} map 
indicates a reddening $E(B-V) = 0.63$ at this position 
($l = 178.08,\, b = -10.31$), and the continuum  
does suggest substantial reddening.  The presence of diffuse 
interstellar
bands near $\lambda 5780$ and $\lambda 6284$ \citep{jenniskensdesert}
gives further evidence for reddening, and 
the rather strong Na D absorption lines
near $\lambda 5893$ are almost motionless in the spectrum,
indicating that they are mostly interstellar.  In computing
the secondary-based distance, we assume that V1062 Tau is
beyond most of the reddening in the \citet{schlegel98} map,
which yields a distance of $\sim 1400$ pc, which would put it
over 200 pc from the midplane of the Galaxy, where it would
indeed be expected to be outside most of the dust layer.

Our spectrum has a synthetic $V$ magnitude near 16.7.  However, we also obtained 
two spectra 2002 Feb 20.53 UT that show the star much brighter,
with synthetic magnitudes 
near $V = 14.7$; another pair of spectra taken at Feb 20.78 UT,
a few hours later, have $V = 14.9$ and 15.3.
By a remarkable coincidence, \citet{lipkin04} observed the 
same outburst earlier in the same night (from a different
longitude); our first pair of spectra was taken some 4.5
hours after their last point.  If the decline between our 
spectra was the decline from outburst, the flare's duration
was considerably less than one day.
Only a few magnetic CVs have been observed
to undergo such short-lived flares \citep{lipkin04}.

In our spectra 
the velocity of the late-type feature is modulated at 
$0.415883(13)$ d, or 9.9812(3) h, consistent with 
the \citet{lipkin04} period.  
The emission-line velocities
include some points from earlier observing seasons; the
signal-to-noise of these spectra was not adequate for measuring
absorption-line velocities, but the increased time base does
result in a more precise period.  The emission-line velocities 
do not determine the period on their own -- the single best fit to
the emission is near 0.4187 d (10.05 h).
However, when the absorption line velocities are fitted at the
10.05 h period, the fit is very poor, while  
the fit to the emission-line velocities at 
9.9812 h (the best absorption-line
period) is reasonably good.  
Because the absorption lines arise in a relatively 
uncomplicated stellar photosphere, their velocities should 
be much less vulnerable to state changes (and the like) than those
of the emission lines; because of this, and the good agreement with
the independent period determination by \citet{lipkin04}, we are 
confident that the absorption-line period is essentially correct.
In Table~\ref{tab:parameters} and Fig.~\ref{fig:folpl1} we adopt 
the weighted average of the emission- and absorption-line periods.
The two periods do disagree by 3.6 standard deviations, so 
some caution is in order extending the ephemeris into the future,
but the actual difference of the periods is only 4.3 s.

The spectral decomposition shows the secondary to be of type
K5 $\pm 2$ subclasses.  At this orbital period and spectral type, 
modeling suggests that CV secondaries depleted much of their 
core hydrogen before the beginning of mass transfer 
\citep{bk00,kniggedonor,podsiadlowski}.

The emission-line velocity modulation is far from antiphased
with the absorption (see Fig.~\ref{fig:folpl1}), so it is
evidently not a faithful tracer of the white dwarf motion. 
In contrast, the absorption lines should follow the secondary
star.  Without a reliable $K_1$ we cannot make firm mass
estimates.  However, the absorption velocity amplitude $K_2$, 
$202 \pm 13$ km s$^{-1}$, is fairly large for this $P_{\rm orb}$,
and suggests that the white dwarf is relatively massive and that
the inclination is not too far from edge-on.  As an illustration,
a system $M_1$ = 1 M$_{\odot}$ and $M_2$ = 0.5 M$_{\odot}$ at
$i = 68$ degree would produce the observed $K_2$.  Note that
\citet{lipkin04} do not detect eclipses despite intensive coverage.

%

\subsection{V344 Ori}

Our spectrum appears similar to that published by 
\citet{munari97}.  We clearly detect an M-dwarf
secondary.  The gross period (Table~\ref{tab:parameters}) is 
based on emission-line
velocities.  It is unambiguous,
but the fine period is aliased on a scale of one
cycle per 47 d due to uncertain cycle count between 
observing runs; the allowed precise periods satisfy
$$P = 47.305 \pm 0.002\ {\rm d} \over 205 \pm 2,$$ where the 
denominator is an integer.
The cross-correlation technique 
was sensitive enough to measure M-dwarf velocities for some
of our spectra, but these were not extensive or accurate enough
to shed light on the period; however, they did confirm that
the M-dwarf light is from the binary companion and not from 
an interloper.  Combining our spectra into a 
single-trailed greyscale representation confirmed the presence
of secondary features; in particular, the CaI $\lambda$ 6122
absorption line appeared faintly, with a velocity semiamplitude of 
$\sim 170$ km s$^{-1}$.

V344 Ori is at $l = 195^{\circ}.05, b = -0^{\circ}.72$, for which
\citet{schlegel98} give $E(B-V) = 1.13$ to the edge of the Galaxy.
However, its spectrum shows neither diffuse interstellar
bands nor interstellar NaD absorption, which would be expected if the
reddening were this large.  Standardized 
$UBVI$ observations taken 2008 Jan.~16.0 UT give 
$V = 18.76$, $B-V = 0.60$, $U-B = -1.01$, and $V - I = 1.72$.
At minimum light, dwarf novae typically have  $B - V \sim +0.1$ and
$U - B \sim -0.8$ (see, e.g., \citealt{vogt83}); the very
blue $U-B$ suggests near-zero reddening, while the redder
$B-V$ suggests $E(B-V) \sim 0.5$.  
The poorly-constrained reddening contributes substantially
to the distance uncertainty (Table~\ref{tab:inferences}).

\subsection{VZ Sex = RX J0944.5+0357} 

\citet{jiangvzsex} obtained a spectrum of this ROSAT all-sky
survey source that showed the broad emission lines of a CV.
\citet{mennickent02} obtained more extensive spectrophotometry
from which they deduced a spectral type of M2 for the secondary.
They also found a radial-velocity periodicity that corroborated an
early version of the period reported here.

Nearly all our observations are from 2001 March, but we also
have two velocities from 2000 April.  The period measured in 2001 March is
not precise enough to extrapolate the cycle count to the previous
year, but the extra data do constrain the precise period to 
a group of aliases which can be expressed as 
$$P = {352.921 \pm 0.013\ {\rm d} \over 2373 \pm 18},$$ 
where the denominator is an integer.

The M-dwarf contribution to the spectrum was strong enough that
we could measure absorption velocities for the majority
of our spectra by cross-correlating the M-dwarf radial velocity 
template over the 6000 to 6500 \AA\ region.  The absorption
velocities were not precise enough to yield an accurate 
$K_2$, but they did move approximately in antiphase to the 
emission, confirming that the M dwarf contribution arises from the 
binary companion.  

We also estimated the secondary star's spectral type, and found
the best match around M1 $\pm$ 1 subtype, slightly earlier than
the type found by \citet{mennickent02}.  This is just a little
warmer than is typical at this $P_{\rm orb}$ (3.57 hr); the 
\citet{bk00} models that predict this early spectral type
begin with relatively massive secondaries and have depleted nearly all 
the core hydrogen at the start of mass transfer.

\subsection{NSVS 1057564+092315}

\citet{greaves} identified this object 
as a CV, most likely a dwarf nova of the UGSS subclass, 
after Patrick Wils first 
called attention to the nature of its light curve from the 
Northern Sky Variability Survey.  It appears in the Sloan
Digital Sky Survey, but was not selected for a spectrum,  
probably because its colors ($g = 15.88$, $u - g = +0.65$, 
$g - r = 0.67$, $r - i = +0.30$, and $i - z = +0.16$) place
it near the `stellar locus' used to exclude spectroscopic
targets \citep{richards02, szkody02}.  This illustrates 
that while the SDSS has been extraordinarily effective at
finding new CVs and illuminating their properties 
\citep{gaensicke09}, it does not find all the CVs within its
footprint (and indeed, \citealt{szkody09} warn that SDSS
does not claim completeness for CVs). 

Our spectra show modulation of both the emission and absorption 
at essentially the same period, with a secure cycle count 
on all timescales over the 1456 d span of our data set.
The weighted average $P_{\rm orb}$ is 
0.376308(1) d (9.03 hr).  
If we define orbital phase $\phi = 0$ at the blue-to-red
crossing of the absorption velocities through their mean
velocity $\gamma_{\rm abs}$, then the emission velocities
cross blue-to-red at $\phi = 0.479 \pm 0.004$, which is 
close to the antiphased behavior expected 
if the emission lines trace
the white-dwarf velocity.   If the emission lines do in
fact trace the white-dwarf velocity (which
is far from proven),  $q = M_2 / M_1 = K_1 / K_2$, 
yielding $q = 0.85(5)$.  
From \citet{bk00}, we find $M_2 = 0.6(1) M_\odot$ for a K4 
or K5 secondary star with a $P_{\rm orb}$ of roughly 9 h; 
this gives $M_1 = 0.7(1)$ from $q$.  These values 
of $P_{\rm orb}$, $M_1$ and $q$ would imply an inclination
near 45 degrees, so eclipses are not expected.

At $l = 241^{\circ}.1$, $b = +57^{\circ}.9$, and $\sim 900$ pc 
(Table~\ref{tab:inferences}), this object is at least several 
hundred pc away from the Galactic plane, so it may belong to the
thick disk.  The secondary's spectrum does not appear unusual,
so it is probably not a low-metallicity halo object.  It 
does not have a significant proper motion listed in the USNOB1.0
\citep{monetusnob} 
or UCAC3 \citep{ucac3} catalogs.

\subsection{V478 Her}

The General Catalog of Variable Stars \citep{gcvs} lists
this star as having variability type IS, that is, showing
rapid irregular variability.  \citet{rittercat} assign 
a superhump period of 0.12 d (2.88 hr), which would imply 
that $P_{\rm orb}$ is slightly shorter than that, which
would put V478 Her in the (roughly) 2- to 3-hour `period gap'.  
The source for the 2.88-hr  period is a photometric 
time series, 3.7 hours in duration, taken by 
T. Vanmunster on 2001 June 7 UT, during an outburst\footnote{The
observation is described at 
http://www.cbabelgium.com/}. This observation showed a 
variation that Vanmunster interpreted as a probable superhump;
on this basis, 
the star is listed in the final edition of the \citet{downesfinal}
CV catalog as an SU-UMa type dwarf nova.  
We were therefore
puzzled when our initial observations showed no sign of a
short-period radial velocity modulation, despite sufficient
signal-to-noise in the H$\alpha$ emission lines to make
such a modulation apparent if it were present.  

More extensive observations extending over several 
seasons indicate that $P_{\rm orb}$ is much longer 
than had been thought, over 15 h.
The secondary spectrum is just detectable, and the
radial velocities of both the 
the H$\alpha$ emission line and the 
absorption spectrum show the long period.
The short period appears to be mistaken, and the 
star has therefore been misclassified.  Because
of its short duration, the light curve
on Vanmunster's site shows only a single minimum, and 
we note that he characterized his detection of superhumps as
``likely'', rather than definitive.
We do not have time series photometry,
and so cannot comment further on the 
cause of the variation he detected.

The emission-line velocities we measure do not fit a sine curve
accurately, and are not in antiphase to the absorption velocities.
The slightly double-peaked structure visible in Fig.~\ref{fig:folpl1}
may be real, or might result from limited sampling.  The 
secondary-star radial velocity amplitude, $K_2 = 138 \pm 10$ km s$^{-1}$,
does not suggest that the masses are out of the ordinary; for
example, a 0.8 M$_{\odot}$ white dwarf and a 0.7  M$_{\odot}$
secondary would produce the observed $K_2$ at $i = 65^{\circ}$.

The orbital period of V478 Her is unusually long for a CV -- it
lies well outside the range tabulated by \citet{bk00}, so the 
secondary-star mass range given in Table~\ref{tab:inferences} is
even more of a guess than usual.  Even so, the long orbital
period (hence large Roche lobe), warm secondary, and faint apparent 
magnitude indicate that V478 Her very distant for a CV; 
at $l = 46^{\circ}.1$ and $b = 29^{\circ}.7$, it is also at 
least 1 kpc from the Galactic plane.  

\subsection{V1059 Sgr}

This is an old nova, having outburst in 1898.  \citet{duerbeck87}
describe the spectrum as having a noisy flat continuum, with 
Balmer and HeII emission.  The continuum in our spectrum is 
indeed fairly flat, but also shows weak broad undulations that 
superficially suggest a companion's contribution.  However,
we could not generate a convincing spectral decomposition with our
library spectra, so these might not indicate a companion.
In part because of the unfavorable declination, we had to observe this
target on several observing runs before we were certain about the
gross period.  This led to an uncertain cycle count 
between runs, which in turn created a complex structure of 
fine-scale aliases.  The 6.87 h
period given in Table~\ref{tab:parameters} is based on the
velocities from our most extensive single observing run.

\subsection{V1082 Sgr}

\citet{cieslinski98} found that the V1082 Sgr shows high
and low states; when it is in a high photometric state ($V \sim$ 14.0 - 14.5), 
it shows emission lines, with HeII $\lambda 4686$ comparable in strength to 
H$\beta$, while in its low state ($V \sim$ 15.0 - 15.5), the emission
lines disappear.  A K-type absorption spectrum is present as well, 
so they suggested that V1082 Sgr may be a symbiotic star. 
The orbital period we find, near 0.868 d, would be very short for a
symbiotic, and is consistent with a long-period novalike
variable.  The Swift satellite detected V1082 Sgr in hard
X-rays (14 -- 195 keV; \citealt{swift22month}); it is one of 29 
CVs found in the Swift BAT 22-month survey catalog,
most of which have proven to be magnetic CVs of the DQ Her 
(or Intermediate Polar) subtype \citep{mukai09}.

Our spectra confirm the state behavior noted by \citet{cieslinski98}, but
we never did see the emission lines disappear entirely.  Strong
He II $\lambda 4686$ is often seen in magnetic CVs; this, 
combined with the hard X-ray detection, suggests that this is
indeed a magnetic system.  Time-series measurements in the 
optical and X-ray might reveal a white-dwarf spin period.

It proved difficult to get a good match to the secondary spectrum,
so the spectral type in Table~\ref{tab:inferences} is 
fairly uncertain.  The unusually long period puts it outside the 
realm of the \citet{bk00} models, so a wide range of $M_2$
is assumed in the distance calculation; fortunately,
this has little effect on $R_2$.  The \citet{schlegel98}
map gives $E(B-V) = 0.16$ at this location ($l = 15^{\circ}.9, b = -12^{\circ}.7$);
this is small enough so that it does not contribute significantly to the 
distance uncertainty, which is dominated by the crude
spectral decomposition.

\subsection{FO Aql}

This dwarf nova is among the few studied by \citet{urban06}
that does not have a known orbital period.  

The velocities from 2007 June and July define a period
of 0.21777(4) d, or 5.23 h.  We also have a few velocities
in 2005 and 2006.  The cycle count from year to year is
ambiguous, but the two most likely choices give 
periods of 0.217735(3) d and 0.217856(4) d, which 
differ in frequency by one cycle per 393 d.  Some 
of our spectra were taken above minimum light.  

While an early M-type secondary star spectrum is evident, it did
not yield usable radial velocities.  The spectral type and 
contribution were fairly uncertain, so the inferences in 
Table~\ref{tab:inferences} are fairly crude.  The interstellar
extinction at this position ($l = 35^{\circ}.9,$ $b = -5^{\circ}.5$) 
is fairly sizeable -- \citet{schlegel98} estimate 
$E(B-V) = 0.42$ to the edge of the dust layer -- and the 
summed spectrum does show weak diffuse interstellar bands.
As a result of all these factors, the final distance estimate
is uncertain by a factor of around 2.

\subsection{V587 Lyr}

V587 Lyr, also designated TK 5, was discovered by 
\citet{kryachko01}, who classified it as a dwarf nova based
on the observation of two outbursts on astrograph
plates.  The 6.6-hour orbital period found here is
within the normal range for dwarf novae.  The spectrum 
shows a clear contribution from a K-type secondary, and broad emission
lines that appear rather weaker than those in otherwise similar stars.

All of our usable spectra are from a single observing run in 2006 June.
Because the absorption velocities quickly yielded $P_{\rm orb}$, we
took a comparatively small number of spectra.
We also obtained standardized
$UBVI$ photometry on 2003 June 19.34 and 2006 August 29.20; the
2003 June observation gave $V = 17.13$, $B - V = +0.76$, 
$U - B = -0.35$, and $V - I = +1.11$, and the 2006 August
observation was very similar.   As is typical, 
the uncertainty in the distance estimate (Table~\ref{tab:inferences}) 
is dominated by the uncertainty in the secondary's spectral
type.

\subsection{V792 Cyg}

V792 Cyg was discovered by \citet{millerwachmann}, who found 
several typical U Gem-type outbursts on photographic plates. 
\citet{bruchschimpke92} published a spectrum taken with the source
well above minimum light, and noted a possible secondary-star 
contribution.

We have observations from runs in 2007 July and 2007 August, which
were separated by about 35 days.  The cycle count between those
two runs is ambiguous.  Both the absorption and emission
velocities slightly favor a period of 0.2972(1) d, but periods of 
0.2996(1) and 0.2948(1) d are also possible.  These alternate periods
differ by about 3 minutes from the most likely choice, so the
rough period is well-determined.  The period quoted in 
Table~\ref{tab:parameters} is computed assuming the most likely 
choice of cycle count is correct, but the Monte Carlo simulation
gives a discriminatory power of only $\sim 0.8$ for the alias choice.

The $\sim$ K5 spectral type of the secondary is about as expected
given the 7.1 h period \citep{kniggedonor}.  The mass ratio
indicated by the ratio of the absorption and emission velocity
amplitudes is $M_2/M_1 = 0.9 \pm 0.1$, but the emission line
radial velocity modulation differs from antiphase with the absorption
by $0.031 \pm 0.007$ cycles, so this calculation is somewhat suspect.
For component masses broadly typical of CVs ($M_1 = 0.8$ M$_{\odot}$,
$M2 = 0.7$  M$_{\odot}$), the $K$-velocities indicate an orbital
inclination of $\sim 45^{\circ}$.  The system is therefore 
unlikely to eclipse.

\subsection{V795 Cyg}

This star was mis-identified in earlier editions of the Downes
catalog, but we were able to contribute a correction based on 
our imaging, and the correct object is indicated in  
\citet{downesfinal}.  The identification
images show the star at $V = 18.40$, $B - V = +0.21$, 
$U-B = -1.15$, and $V - I = +1.58$.  

The radial velocities are from two observing runs only $\sim 20$ d
apart, followed by another several months later.  The choice of cycle
count between these runs is fairly secure, so the 4.35 h period 
(Table~\ref{tab:parameters}) is 
nearly unambiguous, but it is marginally possible that the true 
orbital frequency is higher by $\sim 1/20$ cycle d${-1}$ (corresponding
to a period $\sim 2.4$ min shorter).

The \citet{schlegel98} maps give a total $E(B-V) = 0.30$ in this direction
($l = 65^{\circ}.7$, $b = +5^{\circ}.5$).  We found that the
secondary spectrum fit was easiest when we de-reddened the spectrum
by this amount.  This is somewhat puzzling, since our 
measured colors (especially $U-B$) are typical 
for an unreddened dwarf nova.  In view of this we
assume a generous uncertainty in the reddening when 
computing the secondary-based distance (Table~\ref{tab:inferences}).

\subsection{V811 Cyg}

Our averaged spectrum resembles that obtained by
\citet{munari97}, who described it as a
``text-book example of a CV spectrum".   The 3.76-hour
period is short compared to most of the objects studied
here, and we do not detect the secondary star's contribution.
Most of our data are from observing runs in 2007 June and 
2007 July, only $\sim 16$ d apart, and the cycle count between these
runs is secure; we also have a few observations from
2005.  The cycle count between 2005 and 2007 is unknown, but
the precise periods allowed by the data are well-fitted by
$$P_{\rm orb} = {733.931 \pm 0.010 \ {\rm d} \over 4687 \pm 4},$$
where the denominator is an integer.

\subsection{V542 Cyg}

\citet{bruchschimpke92} present a spectrum of this star,
which they describe as `extremely faint' at the time of their
observations; they did not convincingly detect emission
lines, and pronounced the dwarf-nova classification as
`doubtful' on the basis of the spectrum.  Our spectrum appears 
much more robust, with strong, broad Balmer lines
and the 4.36-hour orbital period is typical of dwarf novae, so 
we confirm that V542 Cyg is indeed a CV.  We have one set of direct
images taken 2005 Sep.~15.20 UT with the Hiltner 2.4 m telescope,
and calibrated with observations of \citet{landolt92} standard stars;
these gave $V = 17.43$, $B - V = +0.74,$ $V - I = +1.54$, and $U - B = -0.80$.
For comparison, the flux-calibrated spectrum 
published by \citet{bruchschimpke92} 
implies $V > 20$, so it seems possible that they
did not have the star in the slit (they found some of their
fainter targets by blind offset).  


There is an obvious late-type contribution in our
spectrum.  The late-type contribution appears to explain the
coexistence of the blue $U-B$ color with the relatively 
red $B-V$ and $V-I$ colors. 
However, the late-type component does not
show a significant radial-velocity modulation at 
the period derived from the emission lines (see Fig.~\ref{fig:folpl2}), 
so the late-type
spectrum clearly is not from the secondary. 
The late-type spectrum could be from an interloper 
elsewhere along the line of sight, but
in our best direct image (in the $I$-band), the point-spread
function has a FWHM of 1.0 arcsec (at 0.275 arcsec pixel$^{-1}$),
yet V542 Cyg shows no sign of incipient resolution.  In addition, the 
position of the blended image does not shift between the images
in different filters, with an upper limit of $\sim 0.15$ arcsec.
The $U$-filter image should be dominated by the CV, while the 
$I$-filter light is dominated by the late-type contribution, so the 
lack of a shift means that the two objects are co-located in the 
sky to good accuracy.  This suggests that the 
late-type contributor is physically associated with the CV, even if 
it is not the Roche-filling companion.  V542 Cyg would seem to be 
a triple system.  
 
\subsection{PQ Aql}

V. Goranskij
\footnote{VSNET-alert number 2140, 1998; http://www.kusastro.kyoto-u.ac.jp/vsnet/Mail/alert2000/msg00140.html}
clarified the identification of this star; he 
noted its blue
color and that it appears to be in outburst
on the first Palomar Observatory Sky Survey, and suggested that
it may be a cataclysmic binary.  
Our spectra show the typical dwarf nova emission lines, confirming
his suggestion.  Nearly all of our spectra were taken in 
quiesence; they are from two closely-spaced observing runs 
runs in 2007 June and July.   The H$\alpha$ emission-line radial velocities
give a best period of 0.24675 d; a much less likely choice
of cycle count between the observing runs yields 0.2436 d.
In some of the spectra we were able to measure the 
velocity of the secondary star.  The absorption velocities
were less accurate than the emission velocities, but they
independently gave the same orbital period.  The secondary's
spectral type, M1.5 $\pm$ 1.0, is approximately as expected
for the 5.9-h orbital period \citep{kniggedonor}.

\subsection{V516 Cyg}

\citet{bruchschimpke92} reviewed the discovery history of 
this star and published a spectrum that
shows it somewhat brighter and bluer than our mean
spectrum.  They described their spectrum as resembling
that of a dwarf nova on the decline from outburst, while
ours appears to represent a more nearly quiescent state.
The continuum shows weak, broad features that resemble
the absorption bands of a secondary star, but 
we were unable to match these convincingly to our 
spectral-type standards, so they may not arise from a
secondary.  Our data come from two short runs separated 
by two years; the 4.1 h period is present in both data
sets, and is definitive in the combined data, but the 
cycle count between runs is not determined.

\subsection{VZ Aqr}

Our spectrum resembles the one obtained by \citet{szkodyhowell92},
except that the flux level in our spectrum is somewhat lower and the 
emission equivalent width appears to be greater.  Like them, we
detect an M-type companion; they estimated the spectral type
to be in the range M0 to M5, and from that estimated the 
distance to be $> 110$ pc, for their latest possible spectral
type.  They did not have a $P_{\rm orb}$ with which
to constrain the size of the Roche lobe.  Our 3.85 h period
is based on H$\alpha$ radial velocities from a single observing
run in 2005 September.

We obtained photometry in 2005 September that showed the
star near $V = 18.0$ on two separate nights; the synthesized
$V$ magnitude from our mean spectrum is also near 18.0.  
Using the secondary star spectrum (Table~\ref{tab:inferences})
we find VZ Aqr to be considerably greater than the minimum
distance found by \citet{szkodyhowell92}.  
\citet{warner87} found an empirical relationship between
the absolute visual magnitudes of dwarf novae at maximum
light and their orbital periods.  This predicts $M_V = +3.8$ at
VZ Aqr's period.  VZ Aqr reaches $V = 11.3$, which would
imply a distance near 320 pc.  Given this, our 
estimate of $600 (+350, -220)$ pc is probably somewhat greater
than the true distance.  As noted above, the normalization
of the spectrum used for the distance calculation is 
corroborated by the photometry derived from direct imaging.
The largest uncertainty in the distance calculation is the
spectral classification of the secondary star.  A closer
distance would suggest a spectral type somewhat cooler
than our estimate.


\subsection{RXJ2133.7+5107}

The X-ray source RXJ2133.7+5107 was discovered by the $ROSAT$ 
Galactic Plane Survey as an X-ray source, and subsequently 
identified as a DQ Her-type CV by \citet{motch98}.  \citet{bonnetrx2133} 
presented photometry which revealed a 570 s modulation,
related to the white-dwarf spin period, and spectroscopy
showing a radial-velocity period of 7.193(16) h, or
0.2998(7) d, which is evidently $P_{\rm orb}$.  
Our H$\alpha$ radial
velocities show modulation at 0.297432(3) d; while
this close to \citet{bonnetrx2133} period, it 
differs by 3.4 standard deviations.  The interval between
our first and last observations is 624 days, and there is
no ambiguity in the cycle count over that time.  
\citet{bonnetrx2133} observed the star on two brief runs some six
years apart, and were naturally unable to determine the 
cycle count between their two epochs, so our period should
supersede theirs.  
Our average spectrum (not shown) is very similar in appearance to that 
that that published by \citet{bonnetrx2133}.  The only new 
information we present for this star is the improved period.

\section{Conclusions}

Our main goal in observing these stars was to determine their
orbital periods, or, if the periods were already known, to 
refine them.  The objects discussed here were selected to 
be those with longer orbital periods.  At longer periods
the secondary star tends to be more conspicuous than at
shorter periods; in accordance with expectation, we detect
the secondary star's spectrum in most of these objects.
The secondaries' spectral types are approximately as expected 
\citep{kniggedonor} for the orbital periods we find.

Most of the present sample appear to be typical long-period
dwarf novae, without obvious abnormalities.  Other types
in the CV zoo are also represented.

Some of our more noteworthy (or at least 
unanticipated) findings are as follows.

\begin{enumerate} 
\item We do not confirm the presence of a white-dwarf
contribution in V709 Cas \citep{bonnetv709} and argue
that it is probably not detectable.
\item IPHAS 0345 resembles a long-period dwarf nova,
strongly dominated by its secondary-star contribution, but
it has never been observed to outburst.  In this respect
it resembles LY UMa.
\item We find that V478 Her is a long-period novalike
variable, rather than a short-period SU UMa-type
dwarf nova.
\item We report the first secondary-star radial velocity
curve for the DQ Her star V1062 Tau, and confirm reports
of flaring activity in the literature.
\item In V542 Cyg, we find a contribution from a late-type
star that is evidently not the secondary star, and argue
that it is likely to be a triple system. 
\end{enumerate}

\section{Acknowledgements}

We gratefully acknowledge support from NSF grants AST-9987334 
and AST-0307413.  We thank the MDM staff for observing 
assistance, and Dartmouth undergraduate Jimmy Chiang for
taking some of the data on V1082 Sgr.  John Southworth 
provided a timely and helpful referee report.  As always, we thank
the Tohono O'odham Nation for letting us use their mountain
for a while.

\clearpage

\begin{deluxetable}{lrrrrrrl}
\tablewidth{0pt}
\tablecolumns{8}
\tablecaption{List of Objects}
\tablehead{
\colhead{Name} &
\colhead{$\alpha_{\rm 2000}$} &
\colhead{$\delta_{\rm 2000}$} &
\colhead{Epoch} &
\colhead{Type} &
\colhead{Maximum} &
\colhead{Minimum} &
\colhead{References} \\
\colhead{} &
\colhead{[h m s]} &
\colhead{[$^{\circ}$ $'$ $''$]} &
\colhead{} &
\colhead{} &
\colhead{[mag]} &
\colhead{[mag]} &
\colhead{} \\
}
\startdata
V709 Cas & 0 28 48.81 & 59 17 22.1 & 2007.7 & DQ & 14.75 B & 15.35 B &  1,2\\
AF Cam & 3 32 15.51 & 58 47 22.4 & 2002.8 & UG & 13.4 p & 17.6 p & 1,2 \\
IPHAS J0345 & 3 45 11.59 & 53 35 14.5 & \nodata & CV & \nodata & 16.0 r$'$ & 3 \\
V1062 Tau & 5 02 27.48 & 24 45 23.2 & 2007.7 & DQ  & 15.9 V & 16.2 V & 1,6\\
V344 Ori &  6 15 18.95 & 15 30 59.3 & 2008.0 & UGZ & 14.2 p & 18.8 V & 1,2,9 \\
VZ Sex   &  9 44 31.72 &  3 58 06.1 & 2000.3 & UG: & 12.8 V& 16.8 V & 1,2,9 \\
NSVS 1057+09 &  10 57 56.30 & 9 23 15.0 & \nodata & UG & 13.0 & 15.5 & 4,8\\
V478 Her & 17 21 05.62 & 23 39 36.8 & 2006.2 & \nodata & 15.5 p& 17.1 p& 1,2\\
V1059 Sgr& 19 01 50.56 & $-$13 09 41.9 & 2003.5 & NA & 4.5 & 17.7 V & 1,2,9 \\
V1082 Sgr& 19 07 21.87 & $-$20 46 50.5 & 2007.7 & \nodata & 14.2 p& 15.9 p& 1,2 \\
FO Aql &   19 16 38.17 &  0 07 37.0 & 2006.4 & UGSS & 13.6 p& 17.5 p& 1,2 \\
V587 Lyr & 19 17 26.47 & 37 10 40.8 & 2006.7 & UG & 14.9 & 17.1 & 1,2,9 \\
V792 Cyg & 19 31 01.03 & 33 47 04.0 & 1992.7 & UGSS & 14.1 & 17.0 & 5,2\\
V795 Cyg & 19 34 34.09 & 31 32 11.9 & 2003.5 & UGSS & 13.4 & 18.4 V& 1,2,9 \\
V811 Cyg & 19 48 23.31 & 36 26 23.3 & 2004.5 & UGSS & 12.7 & $<17.7$ & 1,2\\
V542 Cyg & 19 49 10.51 & 58 31 59.6 & 2005.7 & UGSS & 13.1 & 18.3 & 1,2\\
PQ Aql   & 19 53 06.67 & 12 59 00.8 & 2002.8 & UG: & 13.0 & 15.0 & 1,2\\
V516 Cyg & 20 47 09.81 & 41 55 26.3 & 1992.7 & UGSS & 13.8 p & 16.8p & 5,2 \\
VZ Aqr   & 21 30 24.62 & $-$2 59 17.5 & 2005.7 & UGSS & 11.3 & 18.0 V & 1,2,9\\
RX J2133+51 & 21 33 43.63 & 51 07 25.0 & 2002.8 & DQ & \nodata & 16 B & 1,7\\
\enddata
\tablecomments{Positions, variable star types, and magnitudes for the
stars studied in this paper.  The three numbers given in the
References column are respectively the sources for the celestial position, 
the variability type, and the magnitude range.   
Most of the positions are measured
from direct CCD images taken at MDM, with coordinate solutions
derived from fits to the  UCAC2 \citep{ucac2} or
USNO A2.0 \citep{usnoa2}.  Positions should be accurate to 
$\sim 0''.1$.  Note that the epoch listed
in column 4 is the date the 
image was taken, not the coordinate equinox, which is always
J2000.  References: 
(1) positions from MDM CCD images;
(2) magnitudes from General Catalog of Variable Stars \citep{gcvs}; 
(3) \citet{iphascvs}; 
(4) position transcribed from UCAC-2 \citep{ucac2}; 
(5) position measured from the Digitized Sky Survey served at 
Space Telescope Science Institute.  These are based on scans of 
the Second Palomar Sky Survey, copyright 1993-1995 by the California 
Institute of Technology; 
(6) \citet{remillard94};
(7) \citet{bonnetrx2133}; 
(8) \citet{greaves};
(9) minimum light magnitude from our imaging (see text);
}
\label{tab:star_info}
\end{deluxetable}

\clearpage

\begin{deluxetable}{lrcrc}
\tablewidth{0pt}
\tablecolumns{5}
\tablecaption{Radial Velocities}
\tablehead{
\colhead{Time\tablenotemark{a}} &
\colhead{$v_{\rm abs}$} &
\colhead{$\sigma_{v_{\rm abs}}$}  &
\colhead{$v_{\rm emn}$} &
\colhead{$\sigma_{v_{\rm emn}}$} \\

}
\startdata
\cutinhead{AF Cam:}

51549.8366  & $  136$ & $  21$ & $  -37$ & $  11$ \\
51549.8407  & $   90$ & $  41$ & $  -37$ & $   9$ \\
51549.8447  & $  153$ & $  20$ & $  -30$ & $  10$ \\
51550.8136  & $  101$ & $  22$ & $  -39$ & $   8$ \\
51550.8197  & $  130$ & $  25$ & $  -23$ & $   6$ \\

\enddata
\tablecomments{Full 
table of radial velocities available in the electronic version of this paper.}
\tablenotetext{a}{Heliocentric JD of mid integration 
minus 2,400,000.}
\label{tab:velocities}
\end{deluxetable}

\clearpage

\begin{deluxetable}{lllrrcc}
\tablecolumns{7}
\footnotesize
\tablewidth{0pt}
\tablecaption{Fits to Radial Velocities}
\tablehead{
\colhead{Data set} & 
\colhead{$T_0$\tablenotemark{a}} & 
\colhead{$P$} &
\colhead{$K$} & 
\colhead{$\gamma$} & 
\colhead{$N$} &
\colhead{$\sigma$\tablenotemark{b}}  \\ 
\colhead{} & 
\colhead{} &
\colhead{(d)} & 
\colhead{(km s$^{-1}$)} &
\colhead{(km s$^{-1}$)} & 
\colhead{} &
\colhead{(km s$^{-1}$)} \\
}
\startdata

V709 Cas     & 50801.6342(16) & 0.2222041(3) &  77(4) & $-41(3)$ & 166 &  19 \\[1.2ex]

AF Cam (abs) & 53073.570(4) & 0.3240785(14) &  105(7) & $ 24(5)$ & 36 &  22 \\
AF Cam (emn) & 52233.700(4) & 0.3240780(12) &  63(3) & $ 14(3)$ & 45 &  12 \\
AF Cam (avg) &  & 0.3240782(9) & \\[1.2ex]

IPHAS J0345 (abs) & 54528.802(3) & 0.313902(8) &  77(5) & $ 1(3)$ & 41 &  15 \\
IPHAS J0345 (emn) & 54528.663(4) & 0.313901(11) &  99(7) & $-32(5)$ & 38 &  24 \\ 
IPHAS J0345       & \nodata & 0.313902(7) & \nodata & \nodata & \nodata & \nodata \\[1.2ex]

V1062 Tau (abs) & 53617.988(6) & 0.415883(13) &  202(13) & $ 21(11)$ & 21 &  30 \\
V1062 Tau (emn) & 53450.716(9) & 0.415933(5) &  143(19) & $ 34(13)$ & 27 &  46 \\
V1062 Tau  &\nodata & 0.415926(4) & \nodata & \nodata & \nodata  & \nodata   \\[1.2ex]

V344 Ori & 54525.633(2) & 0.2338(17) &  86(4) & $-27(3)$ & 41 &  15 \\[1.2ex]

VZ Sex (emn) & 51994.6631(12) & 0.1487(3) &  85(5) & $ 59(3)$ & 42 &  16 \\
VZ Sex (abs) & 51994.727(5) & 0.1489(15) &  42(10) & $ 76(7)$ & 33 &  22 \\[1.2ex]

NSVS 1057 (abs) & 54277.7603(16) & 0.3763076(11) &  127(4) & $-3(3)$ & 46 &  12 \\ 
NSVS 1057 (emn) & 54297.509(3) & 0.376312(4) &  107(4) & $-26(3)$ & 54 &  15 \\ 
NSVS 1057  & \nodata  & 0.3763079(11) &  \nodata & \nodata & \nodata &  \nodata \\[1.2ex]

V478 Her (abs) & 53182.904(7) & 0.629049(5) &  138(10) & $ 6(7)$ & 51 &  41 \\ 
V478 Her (emn) & 53182.688(12) & 0.629048(14) &  146(16) & $ 35(12)$ & 135 &  70 \\[1.2ex]

V1059 Sgr (emn) & 54648.932(3) &  0.2861(7)  &  138(11) & $ 49(8)$ & 53 &  36\\[1.2ex]

V1082 Sgr (abs) & 53562.065(12) & 0.867547(19) &  55(5) & $ 47(3)$ & 101 &  17 \\[1.2ex]

FO Aql & 54277.9139(16) & 0.217735(3) &  96(6) & $-33(4)$ & 31 &  12 \\[1.2ex]

V587 Lyr (abs) & 53907.941(4) & 0.2750(10) &  81(7) & $-22(5)$ & 15 &  16 \\
V587 Lyr (emn) & 53907.813(5) & 0.2731(10) &  144(15) & $-22(11)$ & 15 &  32 \\
V587 Lyr & & 0.2740(7) \\[1.2ex]

V792 Cyg (emn) & 54302.618(4) & 0.29715(12) &  118(9) & $-41(7)$ & 28 &  24 \\
V792 Cyg (abs) & 54300.083(2) & 0.29722(19) &  132(8) & $-11(5)$ & 19 &  18 \\ 
V792 Cyg  &   &  0.29717(10)\tablenotemark{c} \\[1.2ex]

V795 Cyg (emn) & 54296.736(2) & 0.181299(17) &  90(6) & $ 13(5)$ & 71 &  21 \\[1.2ex]

V811 Cyg (emn) & 54282.715(2) & 0.15656(4) &  115(13) & $-11(8)$ & 45 &  31 \\[1.2ex]

V542 Cyg (emn) & 53907.738(3) & 0.1815(5) & 88(8) & $-88(6)$ & 20 & 17 \\[1.2ex]

PQ Aql (emn) & 54282.002(3) & 0.24675(4) &  147(9) & $-31(7)$ & 50 &  35 \\ 
PQ Aql (abs) & 54281.885(7) & 0.2470(3) & 155(23) & $ 48(18)$ & 25 &  62 \\
PQ Aql  &  & 0.24675(4) \\[1.2ex]

V516 Cyg (emn) & 54337.0034(13) & 0.1712(4) &  88(5) & $-30(3)$ & 51 &  15 \\[1.2ex]

VZ Aqr (emn) & 53625.867(2) & 0.1606(3) &  44(5) & $-44(3)$ & 63 &  13 \\[1.2ex]

RX 2133 (emn) & 53074.033(3) & 0.297431(5) & 69(5) & $-12(3)$ & 49 & 10 \\[1.2ex]
\enddata
\tablecomments{Parameters of least-squares sinusoid fits to the radial
velocities, of the form $v(t) = \gamma + K \sin(2 \pi(t - T_0)/P$.
Where both emission and absorption velocities are available, the 
period quoted is the weighted average of the periods derived from 
separate fits to the two data sets, and the period is only given
on the first line.}
\tablenotetext{a}{Heliocentric Julian Date minus 2400000.  The epoch is chosen
to be near the center of the time interval covered by the data, and
within one cycle of an actual observation.}
\tablenotetext{b}{RMS residual of the fit.}
\tablenotetext{c}{This period reflects a particular choice of long-term 
cycle count, which remains somewhat ambiguous.  See the text discussion for this 
star.}
\label{tab:parameters}
\end{deluxetable}

\clearpage

\clearpage

\begin{deluxetable}{llrrrrcr}
\tabletypesize{\scriptsize}
\tablewidth{0pt}
\tablecolumns{8}

\tablecaption{Inferences from Secondary Stars}
\tablehead{
\colhead{Star} &
\colhead{Type} &
\colhead{Synthetic $m_V$}  &
\colhead{Assumed $M_2$\tablenotemark{a}} &
\colhead{Deduced $R_2$} &
\colhead{$M_V$\tablenotemark{b}} &
\colhead{$A_V$} &
\colhead{Distance} \\ 
 &
 & 
\colhead{(mag)} &
\colhead{($\rm M_{\odot}$)} &
\colhead{($\rm R_{\odot}$)} &
\colhead{(mag)} &
\colhead{(mag)} &
\colhead{(pc)} \\
}
\startdata
IPHAS J0345 & K5.5 $\pm$ 1.5 & $18.5 \pm 0.5$ & $0.55 \pm 0.20$ & $0.74 \pm 0.10$ & $7.3 \pm 0.5$ & $2.9 \pm 1.3$ & $450(-200,+400)$ \\
V1062 Tau   & K5 $\pm$ 2 & $19.3 \pm 0.4$ & $0.6 \pm 0.15$ & $0.9 \pm 0.1$ & $6.7 \pm 0.7$ & $1.8 \pm 0.5$ & $1400(+700,-500)$ \\
V344 Ori & M3 $\pm$ 0.75 & $20.6 \pm 0.3$ & $0.4 \pm 0.2$  & $0.53 \pm 0.10$ & $10.2 \pm 0.9$ & $0.8 \pm 0.8$ & $600(+500,-270)$ \\
VZ Sex   & M1 $\pm$ 0.75 & $17.9 \pm 0.3$ & $0.28 \pm 0.12$ & $0.35 \pm 0.06$ & $10.2 \pm 0.6$ & $0.1 \pm 0.05$ & $330 (+180, -120)$ \\
NSVS 1057+09 & K4.5 $\pm$ 2& $16.5 \pm 0.3$ & $0.60 \pm 0.15$ & $0.85 \pm 0.08$ & $9.8 \pm 0.7$ & $0.05 \pm 0.03$ & $930 (+360, -260)$ \\
V478 Her & K5 $\pm$ 2 & $19.3 \pm 0.5$ & $0.7 \pm 0.3$ & $1.25 \pm 0.20$ & $6.1 \pm 0.8$ & $0.2 \pm 0.1$ & $4200 (+2100,-1400)$  \\
V1082 Sgr & K4 $\pm$ 0.2 & $16.0 \pm 0.5$ & $0.8 \pm 0.5$ & $1.6 \pm 0.4$ & $5.4 \pm 0.8$ & 
$0.3 \pm 0.2$  & $1150 (+670, -420)$ \\
FO Aql & M2 $\pm$ 1.5 & $19.0 \pm 0.6$ & $0.4 \pm 0.2$ & $0.5 \pm 0.1$ & $10.0 \pm 1.0$ & $0.8 \pm 0.5$ & $440 (+360, -200)$ \\
V587 Lyr & K7 $\pm$ 1.5 & $18.5 \pm 0.3$ & $0.55 \pm 0.20$ & $0.67 \pm 0.09$ & $8.0 \pm 0.7$ & $0.25 \pm 0.15$ & $1120 (+500, -350)$ \\
V792 Cyg & K5 $\pm$ 2   & $18.2 \pm 0.5$ & $0.65 \pm 0.15$ & $0.75 \pm 0.05$ & $7.2 \pm 0.7$ & $0.4 \pm 0.2$ & $1300 (+700, -500)$ \\
V795 Cyg & M3.5 $\pm$ 1.3 & $20.7 \pm 0.6$ & $0.4 \pm 0.2$ & $0.45 \pm 0.08$ & $10.7 \pm 1.1$ & $0.6 \pm 0.3$ & $760 (+620, -340)$ \\
PQ Aql & M1.5 $\pm$ 1.0 & $20.0 \pm 0.8$ & $0.40 \pm 0.15$ & $0.56 \pm 0.08$ & $9.3 \pm 0.6$ & $0.5 \pm 0.2$ & $580 (+190, -190)$ \\
VZ Aqr & M2.0 $\pm$ 1.5 & $19.8 \pm 0.4$ & $0.30 \pm 0.15$ & $0.37 \pm 0.07$ & $8.6 \pm 0.9$ & $0.1 \pm 0.05$ & $600 (+350, -220)$ \\
\enddata
\tablenotetext{a} {Note carefully that these masses are not measured,
but are estimates guided by the models of \citet{bk00}.  They are 
used {\it only} to constrain $R_2$, which depends only on the cube
root of $M_2$, so this does not contribute substantially to the error
budget.}
\tablenotetext{b} {Absolute visual magnitude inferred for the secondary
alone, on the basis of surface brightness and Roche lobe size (see
text.)}
\tablenotetext{c} {The period is above the limit of evolutionary 
scenarios computed by \citet{bk00}, so a conservative estimate of the 
mass is made.}
\label{tab:inferences}
\end{deluxetable}
\clearpage

\clearpage

\clearpage

\begin{figure}
\epsscale{0.91}
\plotone{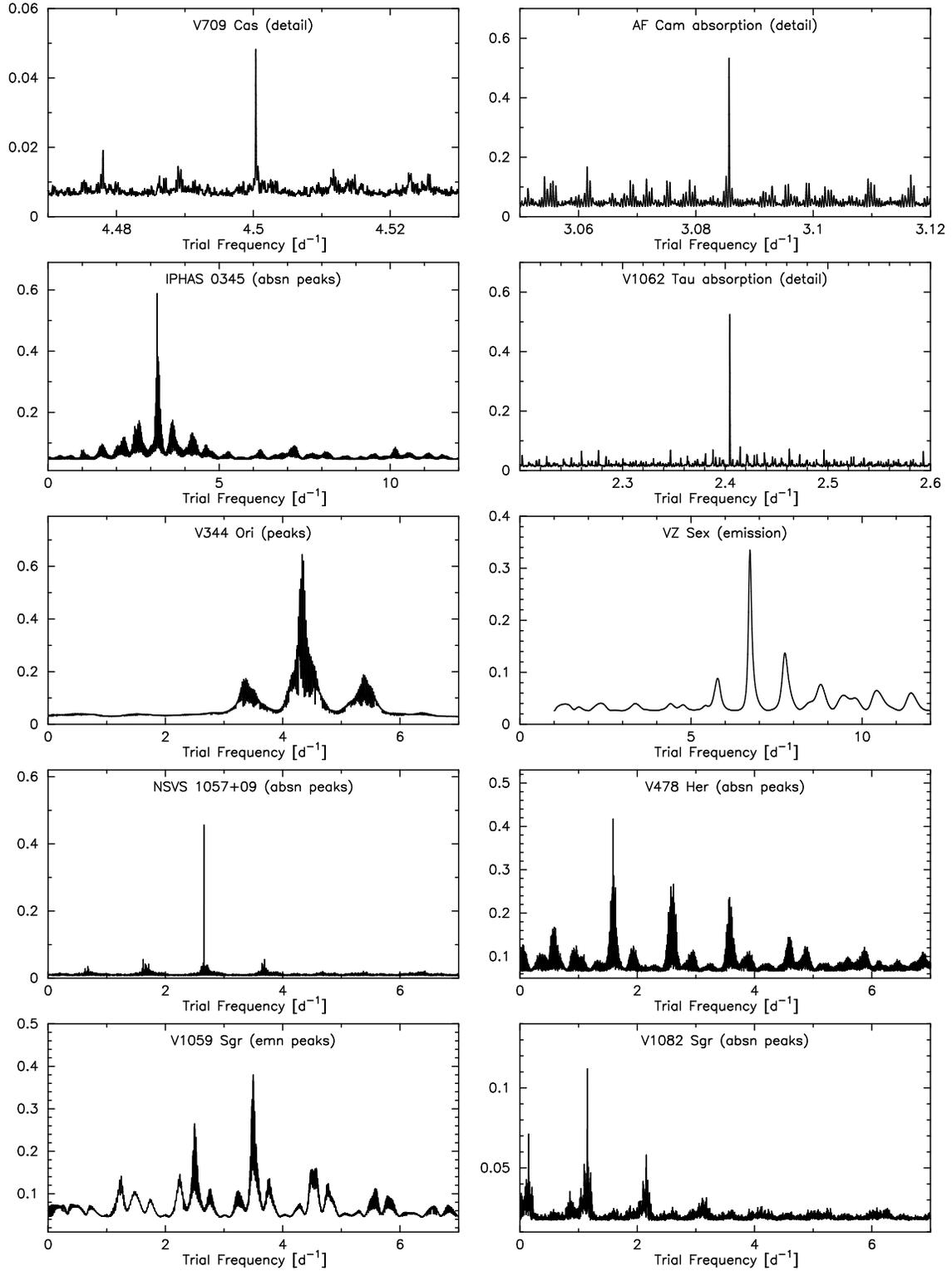}
\caption{Periodograms of the radial-velocity time series, 
computed as described in text.  In those panels labeled as
`peaks', the curve drawn is generated by connecting local
maxima in the full periodogram.  We adopt this expedient in
those cases because the time span covered is long, leading
to an unwieldy number of uninformative fine-scale aliases 
in the full periodogram.
}
\label{fig:pgrams1}
\end{figure}

\clearpage

\begin{figure}
\epsscale{0.91}
\plotone{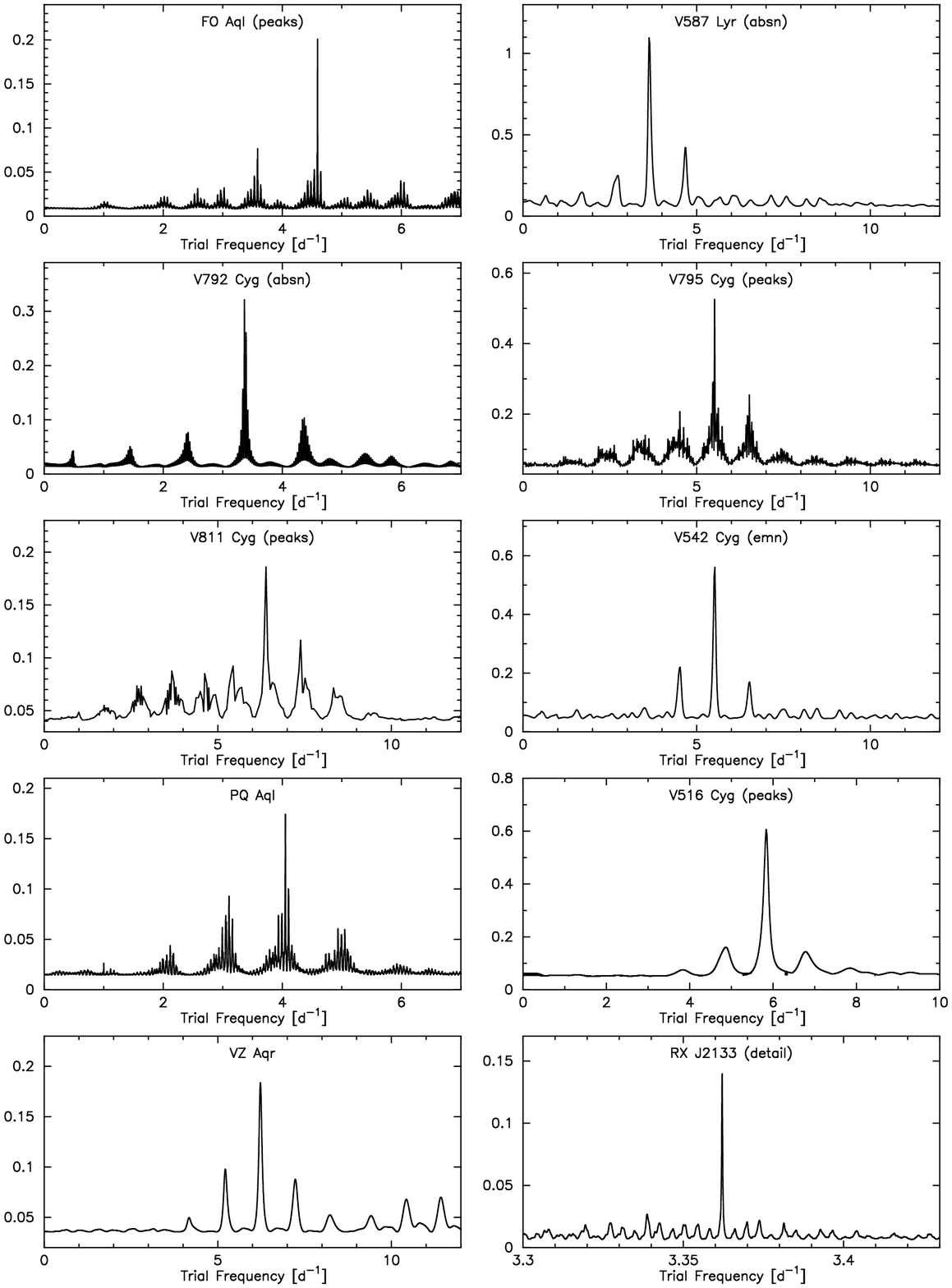}
\caption{Continuation of Fig.~\ref{fig:pgrams1}.
}
\label{fig:pgrams2}
\end{figure}

\clearpage

\begin{figure}
\epsscale{0.91}
\plotone{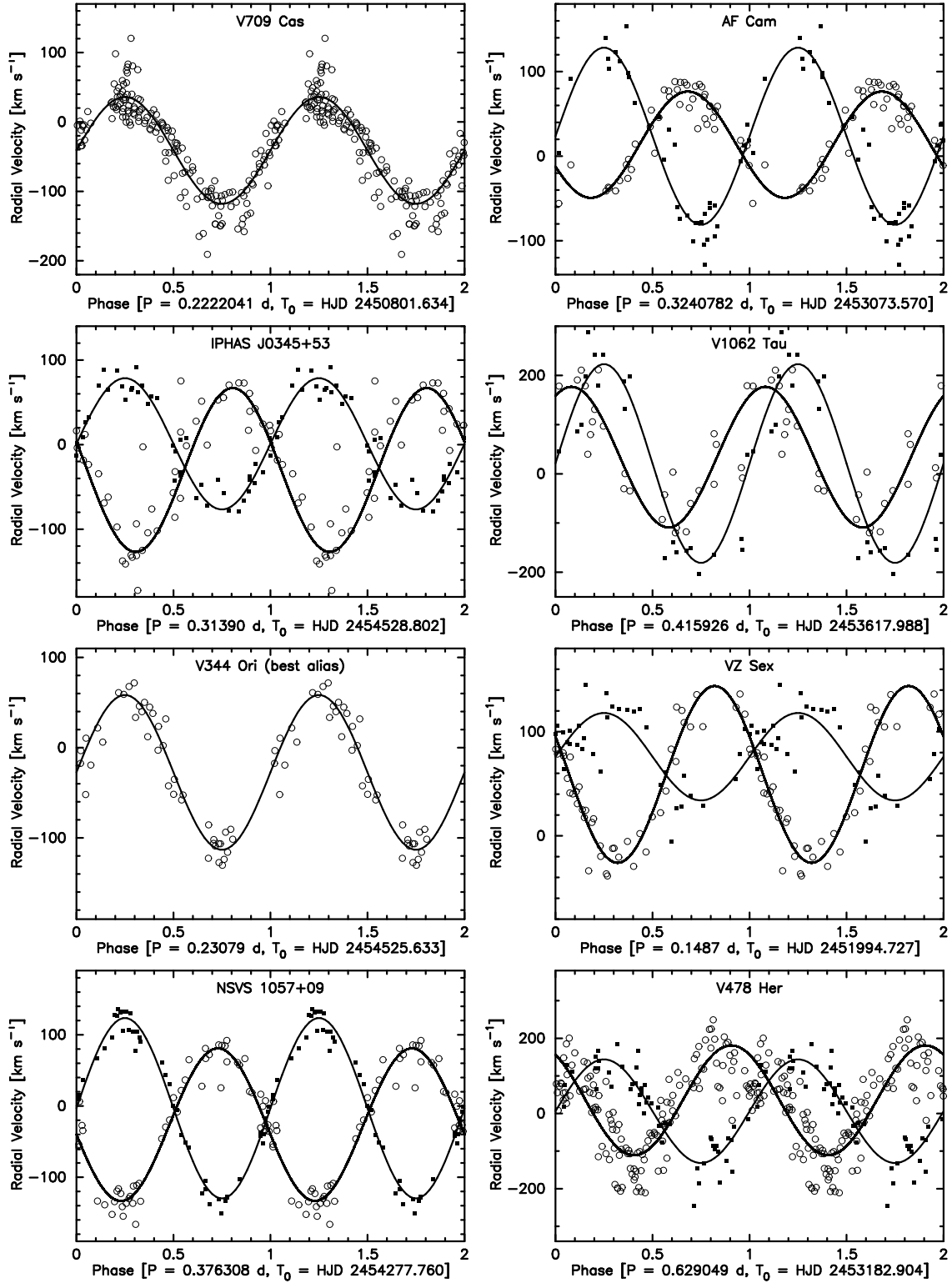}
\caption{Absorption- (solid dots) and emission-line 
(open circles) radial velocities folded on the adopted orbital periods.
For clarity, error bars have been omitted.
Best-fit sinusoids are superposed.  All data are shown twice
for continuity.}
\label{fig:folpl1}
\end{figure}

\clearpage

\begin{figure}
\epsscale{0.91}
\plotone{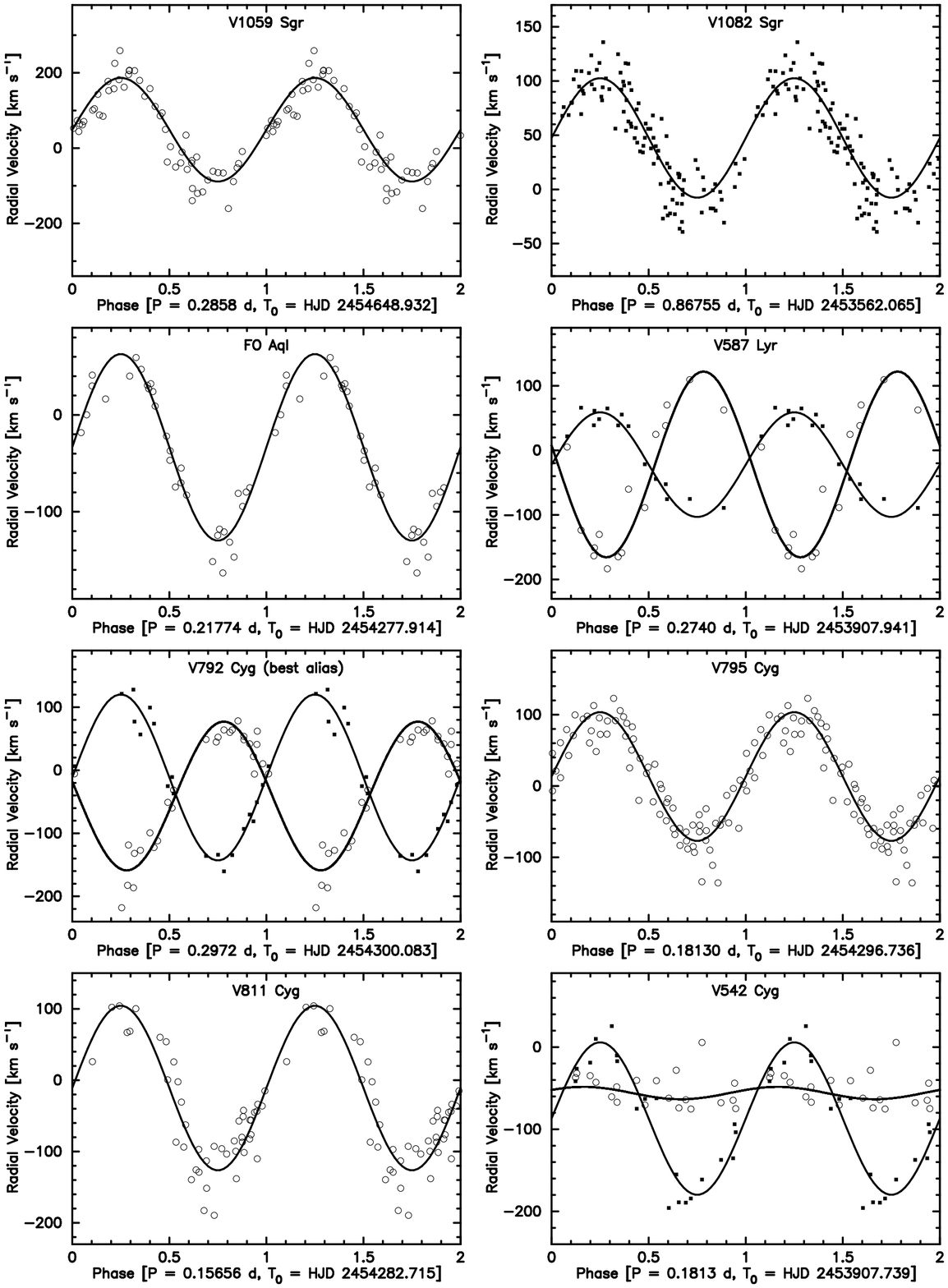}
\caption{Continuation of Fig.~\ref{fig:folpl1}}
\label{fig:folpl2}
\end{figure}

\clearpage

\begin{figure}
\epsscale{0.91}
\plotone{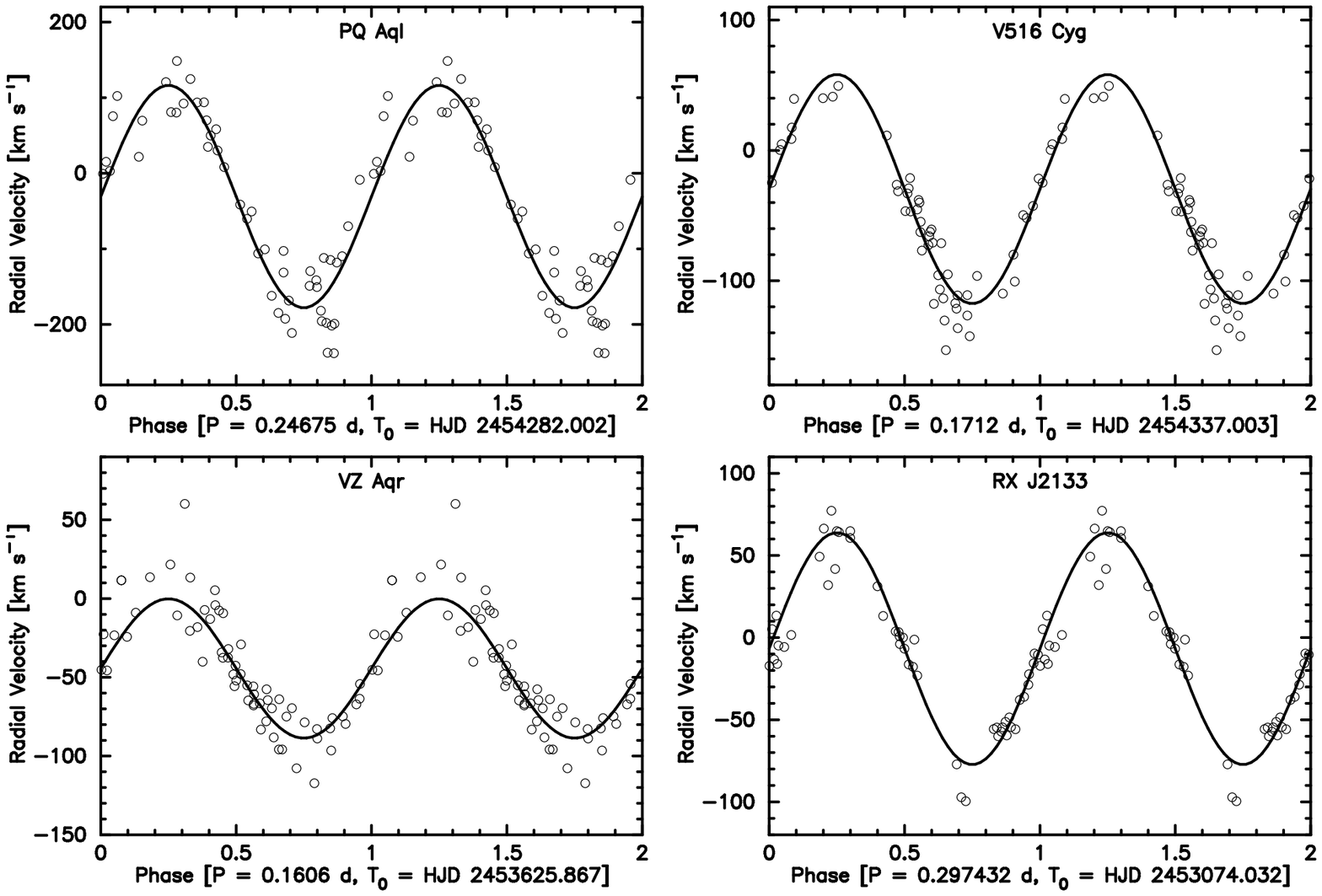}
\caption{Continuation of Fig.~\ref{fig:folpl2}}
\label{fig:folpl3}
\end{figure}

\clearpage

\begin{figure}
\epsscale{0.92}
\plotone{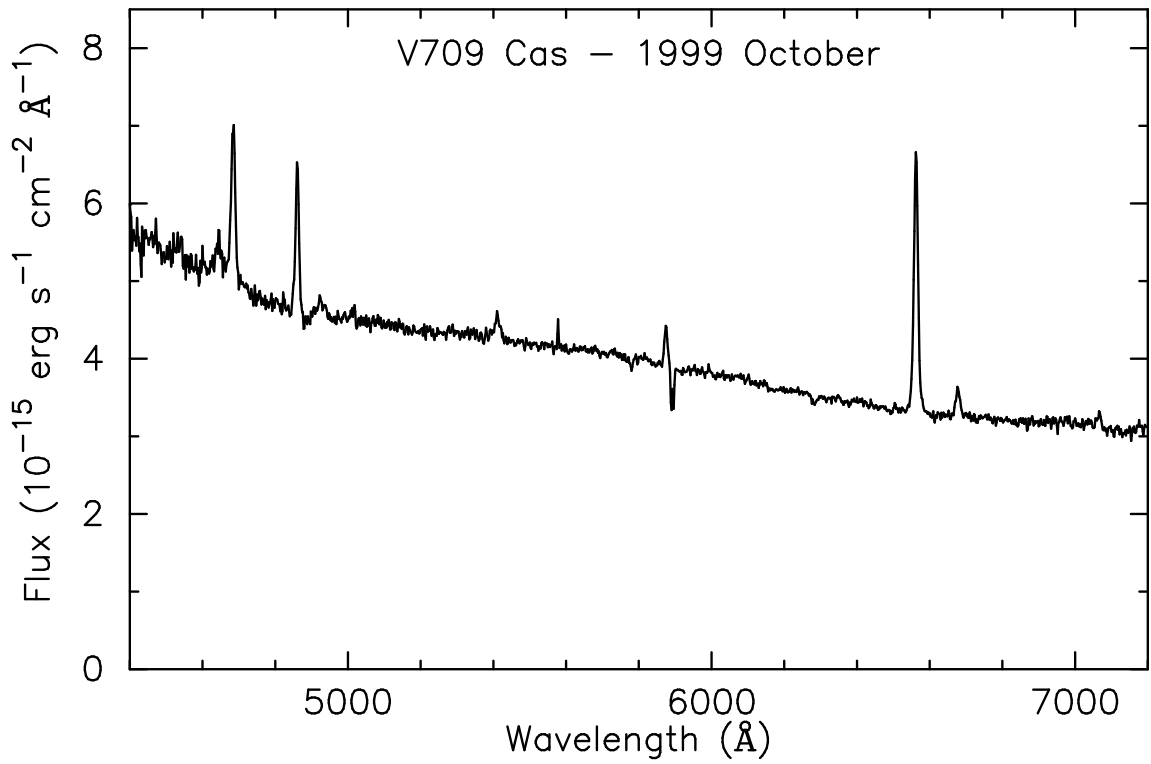}
\caption{Spectrum of V709 Cas from 1999 October.  Note the
lack of white-dwarf absorption features around the Balmer lines, 
and the lack of a discernible late-type contribution. 
The shape of the continuum is not particularly trustworthy.  
}
\label{fig:v709casspec}
\end{figure}

\end{document}